\documentclass[a4paper,superscriptaddress,twocolumn,prl]{revtex4}
\usepackage[pdftex]{graphicx}
\usepackage[]{natbib}
\pdfoutput=1
\usepackage{todonotes}
\usepackage{subfigure}
\usepackage{float}

\newcommand{\ignore}[1]{}

\begin{document}

\title{The Role of Source Coherence in Atom Interferometery}

\author{Kyle S. Hardman}

\email{kyle.hardman@anu.edu.au}
\homepage{http://atomlaser.anu.edu.au/}
\author{Carlos C. N. Kuhn}
\author{Gordon D. McDonald}
\author{John E. Debs}
\author{Shayne Bennetts}
\author{John D. Close}
\author{Nicholas P. Robins}

\affiliation{Quantum Sensors and Atom Laser Laboratory, Department of Quantum Science, Australian National University, Canberra, 0200, Australia}

\date{\today} 

\begin{abstract}
The role of source cloud spatial coherence in a Mach-Zehnder type atom interferometer is experimentally investigated.  The visibility and contrast of a Bose-Einstein condensate (BEC) and three thermal sources with varying spatial coherence are compared as a function of interferometer time.  At short times, the fringe visibility of a BEC source approaches $100$ $\%$ nearly independent of $\pi$ pulse efficiency, while thermal sources have fringe visibilities limited to the mirror efficiency.  More importantly for precision measurement systems, the BEC source maintains interference at interferometer times significantly beyond the thermal source.
\end{abstract}

\maketitle
\subsection{Introduction}
High precision atom interferometers are currently being developed and implemented for an array of practical and fundamental physics applications, including inertial sensing \cite{RobinsAtomLaser}, gravitational wave detection \cite{BramGrav}, measurements of the fine structure constant \cite{Alpha2011}, and tests of general relativity \cite{MullerGravityTest}.  The precision of these devices is proportional to the signal visibility multiplied by the enclosed space time area \cite{PhysRevLett.67.177, Bordé198910, PhysRevLett.42.1103, 2013arXiv1311.2143M}. For example, in a Mach-Zehnder type atom interferometer the minimal acceleration signal that can be measured at the shot noise limit is given by $\delta a=\frac{1}{k_{eff} V\sqrt{N}T^{2}}$, where $V$ is the fringe visibility, $k_{eff}$ is the effective laser wave number, $T$ is the interferometer time, and $N$ is the total atom number.  In order to increase sensitivity it is therefore imperative to maintain fringe visibility at longer interferometer times.  unfortunately, in real systems various factors contribute to a loss in fringe visibility as $T$ is extended including the interferometer cloud's transverse momentum \cite{Louchet-Chauvet:2011aa}, mode mismatch at the recombination stage, and the inability to produce perfect beam splitters and mirrors which introduces impurities into the system.  Traditionally, these problems have been partly mitigated by utilising velocity selection to dramatically narrow the longitudinal momentum width of the interferometer source cloud \cite{0026-1394-38-1-4, Muller24hk, 1367-2630-15-2-023009}.  In such a system, the transverse momentum of the source cloud is unaffected by the velocity selection pulse.  A number of more recent works have utilized ultra-cold sources to narrow both the transverse and longitudinal momentum widths in order to circumvent the decrease in fringe visibility associated with classical effects \cite{80hkarxiv, PhysRevLett.111.083001} \footnote{just as an optical laser gives a {\em practical} advantage over a lightbulb}.

This paper shows that the phase shift and fringe visibility of a Mach-Zehnder atom interferometer (MZI) are critically dependent upon the spatial coherence of the source.  A source cloud with a longer coherence length is shown to produce a more robust atom interferometer.  The comparison of fringe visibility and contrast with $T$ is shown for various cloud coherence lengths down to and below Bose-Einstein condensation.  For short $T$, an increase in fringe visibility is shown for a Bose-Einstein condensate (BEC) as compared to various thermal sources.  Similarly, interference is demonstrated using a BEC, for $T$ long after zero contrast is measured for thermal sources. All clouds used in the interferometer are transversely confined in a horizontal optical waveguide to keep the transverse spatial dimensions constant, and velocity selected to ensure that the longitudinal momentum widths and atom numbers are equivalent.

\subsection{Apparatus}

The apparatus is carefully designed to control the interferometer cloud properties.  Initially, a hot vapor of $^{87}$Rb is produced from an alkali metal dispenser and pre-cooled with a 2D magneto-optical-trap (MOT).  The atoms are then transported through an impedance by a $0.45$ mW push beam, $6$ MHz red detuned from the $\left|F=2, F'=3\right\rangle$ transition, to the secondary cooling 3D MOT stage.  The atoms are polarization gradient cooled to $\approx15\mu$K before being loaded into a hybrid quadupole-magnetic and optical-dipole trap, where they are further cooled to $\geq 1$ $\mu$K using radio frequency evaporation of atoms in the $\left|F=1, m_F=-1\right\rangle$ ground state. These atoms are then loaded into an optical cross dipole trap originating from two separate lasers, one at $\lambda = 1090$ nm, with a $2$ nm linewidth, which is strictly used for the evaporation process and the second at $\lambda = 1064$ nm, with a $1$ MHz linewidth, which doubles as an evaporation and an optical waveguide beam for the interferometer.  The two optical dipole beams are ramped from an initial power of $12$ W each to a set value of $5$ W for the waveguide beam.  Depending on the desired transverse momentum width and phase space of the source cloud, the duration ($3$ and $6$ seconds) and the final power of the main evaporation beam ($2.4$, $2$, and $1.2$ W) is varied.  Immediately following the final evaporation stage, the main evaporation beam is ramped off.  Simultaneously, the waveguide beam is ramped up  to $7$ W over $250$ ms to load the atoms in the waveguide, where they are supported against gravity.  The lifetime of the cloud in the wave guide is on the order of seconds.

 The optical beam splitters and mirror used in this experiment are generated using Bragg transitions.  The Bragg (optical lattice) laser used in this experimental setup has been described previously \cite{PhysRevA.87.013632}.  A maximum power of $100$ $mW$, $105$ $GHz$ detuned to the blue of the $D_{2}$ $\left|F=1\right> \rightarrow\left|F'=2\right>$ transition, is split equally between two counter propagating beams to form the Bragg lattice, which itself is aligned co-linear with the optical waveguide.  The Bragg beams are collimated to a $1.85$ mm $\frac{1}{e^2}$ width at the atoms.  Two independent acoustic optical modulators control the relative frequency of the Bragg beams.

\subsection{Cloud Properties}
 The waveguide confines the source cloud's transverse spatial extent, eliminating expansion in the transverse direction during the interferometer.  This limits visibility decay caused by the sampling of the beam splitter beam's aberrations as the clouds propagate through the interferometer arms, a process that can imprint varying phases dependent on the cloud's spatial position and extent.  The transverse size of the BEC source during the interferometer was calculated to be $\approx 9$ $\mu m$, following \cite{PhysRevA.65.053612}.   Upon release into the guide the initial transverse radius given by $R_{\perp}=\sqrt{2 \mu / M \omega_{\perp}^2}$ decreases due to mean field dissipation .  This is given by  $R_{\perp}(t)=R_{\perp}(1+ \omega_{\parallel}^2 t^2/2-\omega_{\parallel}^4t^4/12)^{-1/4}$, where $\mu$ is the chemical potential, $M$ is the atomic specie's mass, $t$ is the expansion time in the waveguide, $\omega_{\perp}=2\pi \times 70$ Hz is the waveguides transverse frequency and $\omega_{\parallel}=2\pi \times 9$ Hz is the longitudinal trap frequency just prior to loading into the waveguide.  The transverse spatial widths of the thermal sources were calculated by extrapolation from free space ballistic expansion and found to be $54$, $21$, and $13$ $\mu$m for the $2.1$, $0.68$, and $0.41$ $\hbar$k transverse momentum width clouds respectively.  After the waveguide is loaded, an expansion period of $70$ ms is allowed for mapping of longitudinal momentum $\Delta p_{\parallel}$ onto position space and dissipation of the condensate's mean field energy. 
 
\begin{figure}
\centering{}
  \includegraphics[width=1\columnwidth]{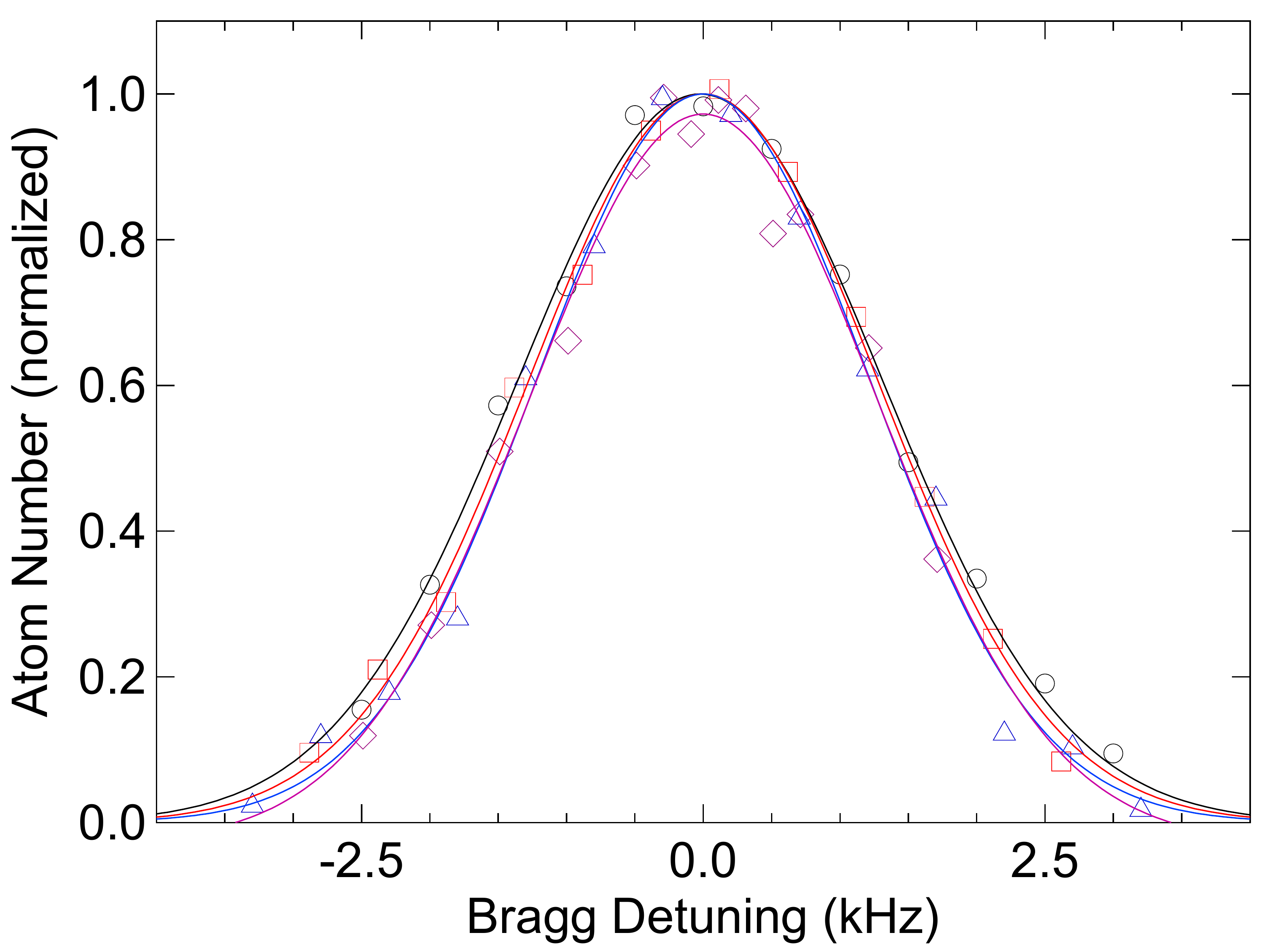}
 \caption{(Color online) Bragg spectroscopy curves of the velocity selected source clouds.    Solid lines are fits to the plotted data points.  Circles: BEC source, $\Delta p_{\perp}<0.2$ $\hbar$k, with fitted Gaussian width $\Delta p_{\parallel}=3.78(8)$ kHz.  Squares:  $\Delta p_{\perp}=0.41$ $\hbar$k thermal source with fitted Gaussian width $\Delta p_{\parallel}=3.6(2)$ kHz.   Triangles: $\Delta p_{\perp}=0.68$ $\hbar$k thermal source with fitted Gaussian width $\Delta p_{\parallel}=3.5(2)$ kHz.   Diamonds: $\Delta p_{\perp}=2.1$ $\hbar$k thermal source with fitted Gaussian width $\Delta p_{\parallel}=3.6(4)$ kHz.  All curves have been normalized in atom number.}
 \label{BraggSpecFig}
 \end{figure}

To ensure consistent longitudinal momentum widths, $\Delta p_{\parallel}$, and atom numbers between differing interferometer clouds, a Bragg velocity selection pulse is used.  Due to the large momentum width of the thermal source clouds, often more than one momentum state is coupled out leading to a large background of impurities in the initial velocity selected cloud.  To separate the desired momentum state from the background, a second velocity selection pulse is used.  Atom number is controlled by adjusting the power of the velocity selection pulse.   There is a $\pm 15$ $\%$ deviation in atom number across all source clouds, with a mean value of $1\times 10^{5}$ atoms.  The longitudinal momentum of all velocity selected clouds is measured by means of Bragg spectroscopy \cite{PhysRevLett.82.4569}.  A $600$ $\mu$s Bragg pulse, out-coupling $\approx 0.032$ $\hbar$k width slices from the interferometer clouds was used.  The $\Delta p_{\parallel}$ of all velocity selected clouds were found to be $\approx 0.12$ $\hbar$k and agree within the measurement uncertainty, as seen in Fig \ref{BraggSpecFig}.  

\subsection{Interferometer Configuration}
There are many configurations of atom interferometers whose sensitivities scale differently with $T$ \cite{PhysRevLett.67.177, Bordé198910, PhysRevLett.42.1103, 2013arXiv1311.2143M}.   An MZI, consisting of a three pulse Bragg sequence, with $T^{2}$ sensitivity dependence \cite{PhysRevLett.42.1103}, is implemented in this experiment.   Using an MZI configuration ideally ensures there is no phase dependence upon the velocity selected cloud's initial velocity.  The velocity selected cloud is first split into two arms through the interaction with a $\frac{\pi}{2}$ Bragg pulse, placing the cloud in a $50/50$ superposition between $0$ and $10$ $\hbar k$  momentum states.  A time, $T$, later, a $\pi$ pulse swaps the momenta of each arm, making the $10$ $\hbar$k states $0$ $\hbar$k and vice versa.  The two arms of the interferometer will be spatially overlapped a time, $T$, later and a final $\frac{\pi}{2}$ pulse is applied to interfere the two momentum states.   After the recombination pulse, the two clouds are given $\approx 25$ ms of expansion in the waveguide to spatially separate.  At this time, the waveguide is switched off and the final states undergo $10$ ms of ballistic expansion before being imaged. 

\begin{figure}
\centering{}
  \includegraphics[width=1\columnwidth]{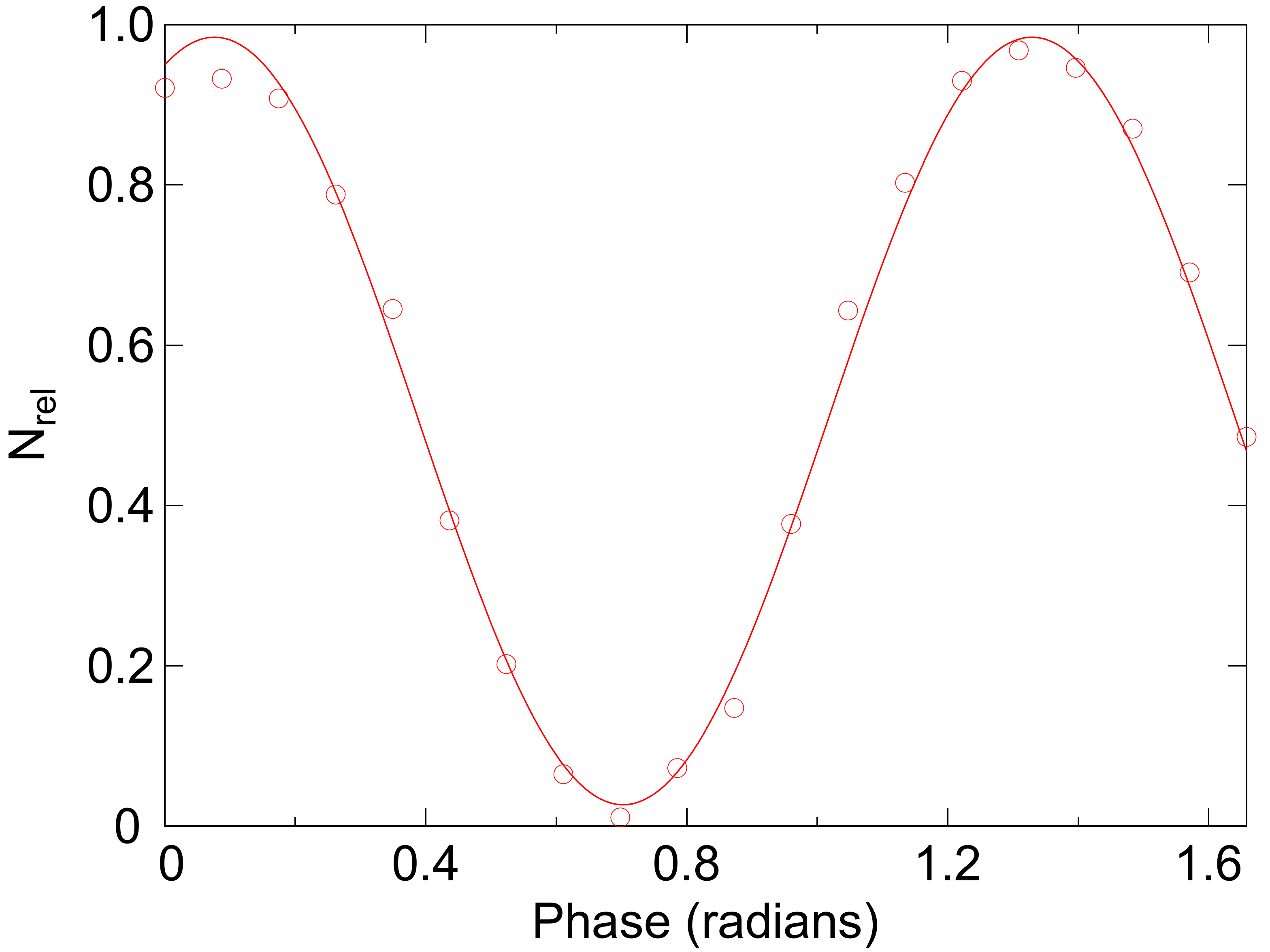}
 \caption{Sample fringe with visibility approaching $100$ $\%$, taken from a $T=0.2$ ms, $10$ $\hbar$k MZI, using a BEC source. This data was taken using a $\pi$ pulse with $\approx 95$ $\%$ efficiency.  Phase along the horizontal axis is adjusted by varying the laser phase of the final ${\frac{\pi}{2}}$ pulse.  The fringe period is $1.25$ radians corresponding to the Bragg order, $n=5$, where fringe period is $\frac{2 \pi}{n}$ radians.}
 \label{Fringe100}
 \end{figure}

The total number of atoms in each final state is counted using absorption imaging on resonance with the $\left|F=2\right> \rightarrow\left|F'=3\right>$ $D_{2}$ transition.  Prior to imaging the atoms are pumped to the $\left|F=2\right>$ ground state with a repump pulse on the $\left|F=1\right> \rightarrow\left|F'=2\right>$ $D_{2}$ transition.  The final number of atoms in state $1$ is normalized to the total number and given by $N_{rel}=\frac{N_{1}}{N_{1}+N_{2}}$, where $N_{1}$ and $N_{2}$ correspond to the total atom number in each respective momentum state.  This is done to eliminate the run-to-run total atom number fluctuations.  The absorption imaging data is analyzed with a Fourier decomposition algorithm, described previously \cite{PhysRevA.87.013632}, to determine which parts of the image contribute to the interference.  By scanning the phase of the recombination pulse, a fringe in relative atom number and phase is observed and shown in Fig \ref{Fringe100}.  The obtained interference fringe oscillates as $N_{rel}=$$y_{0}+V\cos (\phi_{0}+n\phi_{\frac{\pi}{2}})$, where $y_{0}$ is the $N_{rel}$ offset, $V$ is the fringe visibility, $\phi_{0}$ is the phase offset, $n$ is the Bragg order, and $\phi_{\frac{\pi}{2}}$ is the phase of the recombination pulse.

\subsection{Interferometer: $T=0.2 \rightarrow 2$ ms}
It has been shown in previous experiments and reconfirmed in this work, that with an ideal system, ie. near perfect $\frac{\pi}{2}$ and $\pi$ pulses, fringe visibilities approaching $100$ $\%$ can be reached for BEC sources.  This system also exhibits contrast approaching $100$ $\%$ for thermal sources, at short $T$.  In order to investigate the robustness of different sources, a non ideal system was created.  This was accomplished by reducing the amplitude of the $\pi$ pulse until only $80$ $\%$ transfer was achieved.  By producing a non ideal pulse, background impurities are added into the arms of the interferometer.  These impurities are analogous to many practical problems which arise when separating a single cloud into different momentum states.  Impurities such as these can be introduced to interferometer systems in many ways, including scattering into a continuum of momentum states during the cloud separation \cite{Williams20012012, PhysRevLett.93.173201} and the creation of multiple (more than two) states created during Bragg LMT \cite{Szigeti:2012aa} or successive imperfect single Bragg order pulses \cite{Szigeti:2012aa, Kasevich102hk}.  

 A fundamental difference existing between BEC and thermal clouds is the realization of high enough phase space density to allow spatial overlap of the atomic wave functions.  This leads to a spatial coherence for the BEC extending beyond the thermal de Broglie wavelength $\lambda_{T}= \sqrt{2 \pi \hbar^{2}/M k_{b} \mathcal{T}} $, where $\hbar$ is Plank's constant, $k_{b}$ is Boltzmann's and $\mathcal{T}$ is temperature.  In this case the spatial coherence is limited only by the cloud's spatial extent \cite{PhysRevLett.83.3112}.  
 
\begin{figure}
\centering{}
  \includegraphics[width=1\columnwidth]{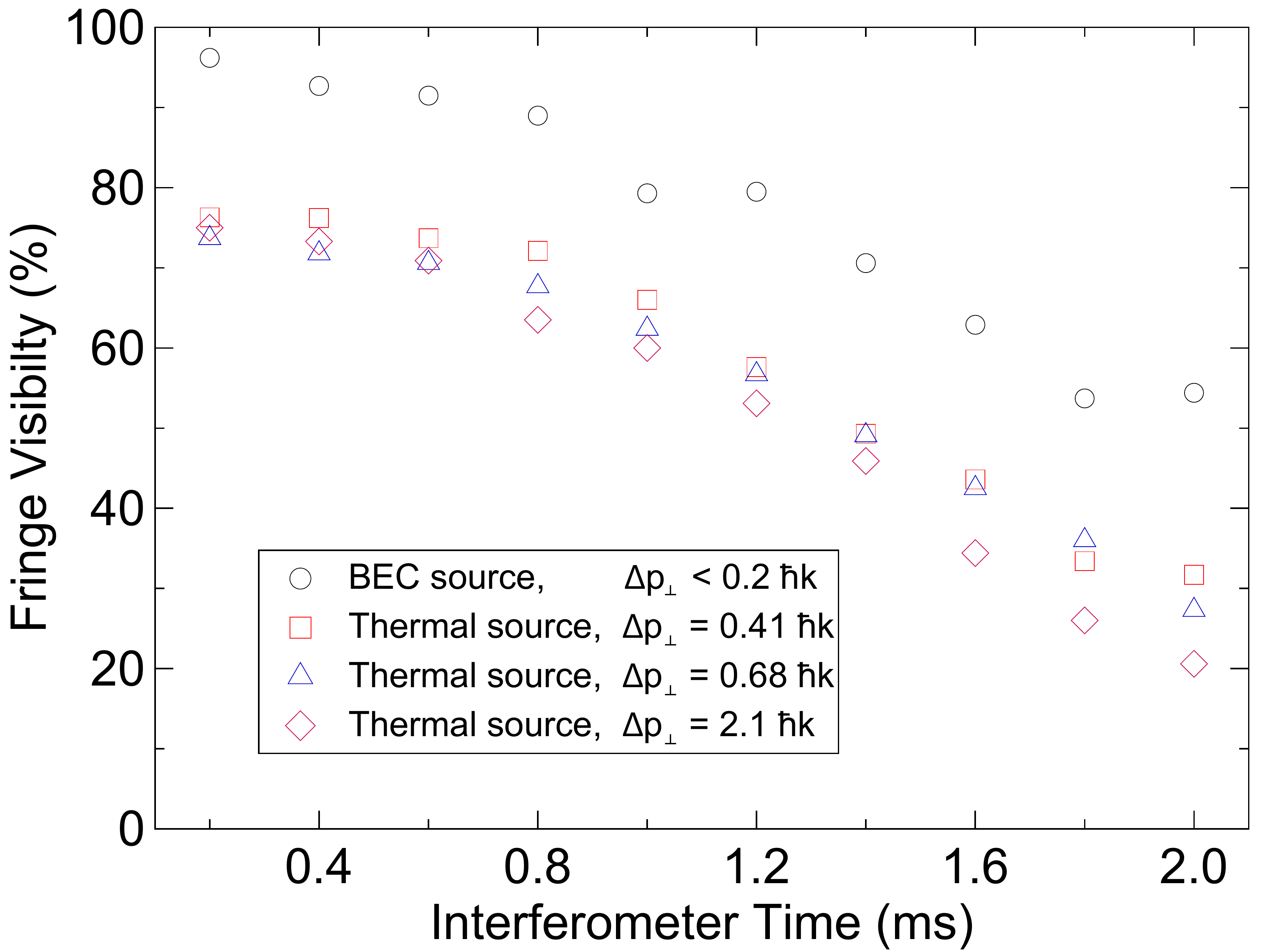}
 \caption{(Color online)  For this data set an $80$ $\%$ $\pi$ pulse was used.  The fringe visibility, as measured from the peak to peak amplitude of a sinusoidal fit, vs interferometer time, $T$, for four separate interferometers with varying levels of source cloud coherence is shown.  Atom numbers between clouds are kept consistent to within $30$ $\%$.  A velocity selection pulse is used to ensure equal longitudinal momentum for all source clouds.  The interferometer is in a Mach-Zehnder configuration consisting of three Bragg pulses, $\frac{\pi}{2}$, $\pi$ and $\frac{\pi}{2}$ with $T$ waited between pulses.  The curves clearly show an increase in fringe visibility beyond the $\pi$ pulse efficiency for the BEC source as well as similar decay rates for all sources.}
 \label{visibility}
 \end{figure}

 This property is seen clearly from the increase in fringe visibility at small $T$ above the limit of the $\pi$ pulse in Fig \ref{visibility}.  Fig (\ref{visibility}) shows fringe visibility as a function of interferometer time, $T$, for three thermal cloud sources with transverse momentum widths of $2.1$, $0.68$ and $0.41$ $\hbar$k and a BEC source.  The $\pi$ pulse efficiencies for all interferometers were adjusted to be $80\pm 2$ $\%$.   The introduced impurity clouds will contribute to the BEC interference signal until the impurities are no longer spatially overlapped with the main clouds during the final $\frac{\pi}{2}$ pulse.  For the system described here, this corresponds to $T=4.5$ ms, which is dependent on the momentum imparted to the clouds during the first $\frac{\pi}{2}$ pulse, $\Delta p_{\parallel}$, and the velocity selected cloud's initial longitudinal size.  At an interferometer time of $T \gtrsim 0.01$ ms, the impurity clouds created from the imperfect $\pi$ pulse are spatially separated beyond the thermal cloud's spatial coherence length which is determined by $\lambda_{T}$ \cite{Bloch:2000yj}.  Beyond this time, the impurity states will no longer interfere with the main thermal interferometer clouds. The impurity clouds then add to the overall signal as a background, which limits the interferometer visibility to the efficiency of the $\pi$ pulse.

The spatial confinement of the source cloud in the waveguide ensures only a small fraction, $<3$ $\%$, of the Bragg beam's wave front is sampled during the interferometer, eliminating the decay associated with the transverse expansion of the cloud.  The decay seen in Fig \ref{visibility} can then be attributed to an effective spatial separation  of the main clouds during the final $\frac{\pi}{2}$ pulse, resulting from another noise source such as transverse oscillations \cite{80hkarxiv} as the clouds propagate.  If the three thermal decay curves were extrapolated to zero visibility, this would correspond to $T$, at which the spatial offset of the clouds at the recombination pulse will have exceeded the thermal de Broglie wavelength, ie. coherence length.  However, the coherence length of a BEC source extends beyond the thermal de Broglie wavelength to the spatial extent of the cloud and therefore fringe visibility will be maintained until a much larger effective spatial offset. 

In the BEC interferometer the spatial mismatch of the four interfering states should exhibit a spatial interference pattern, analogous to an optical four slit experiment.  The fringe spacing is dictated by the atomic system's de Broglie wavelength $\lambda_{S}=\frac{2\pi \hbar}{\Delta p}$, where $\Delta p$ is the momentum difference between the interfering clouds, $T$, and expansion time \cite{PhysRevA.81.043608}.  The spatial interference fringes have a period on the order of microns and are not observed in the final clouds due to the limited resolution of the imaging system, $\approx 30$ $\mu m$.

\subsection{Numerical Modeling}
\begin{figure}
\centering{}
  \includegraphics[width=1\columnwidth]{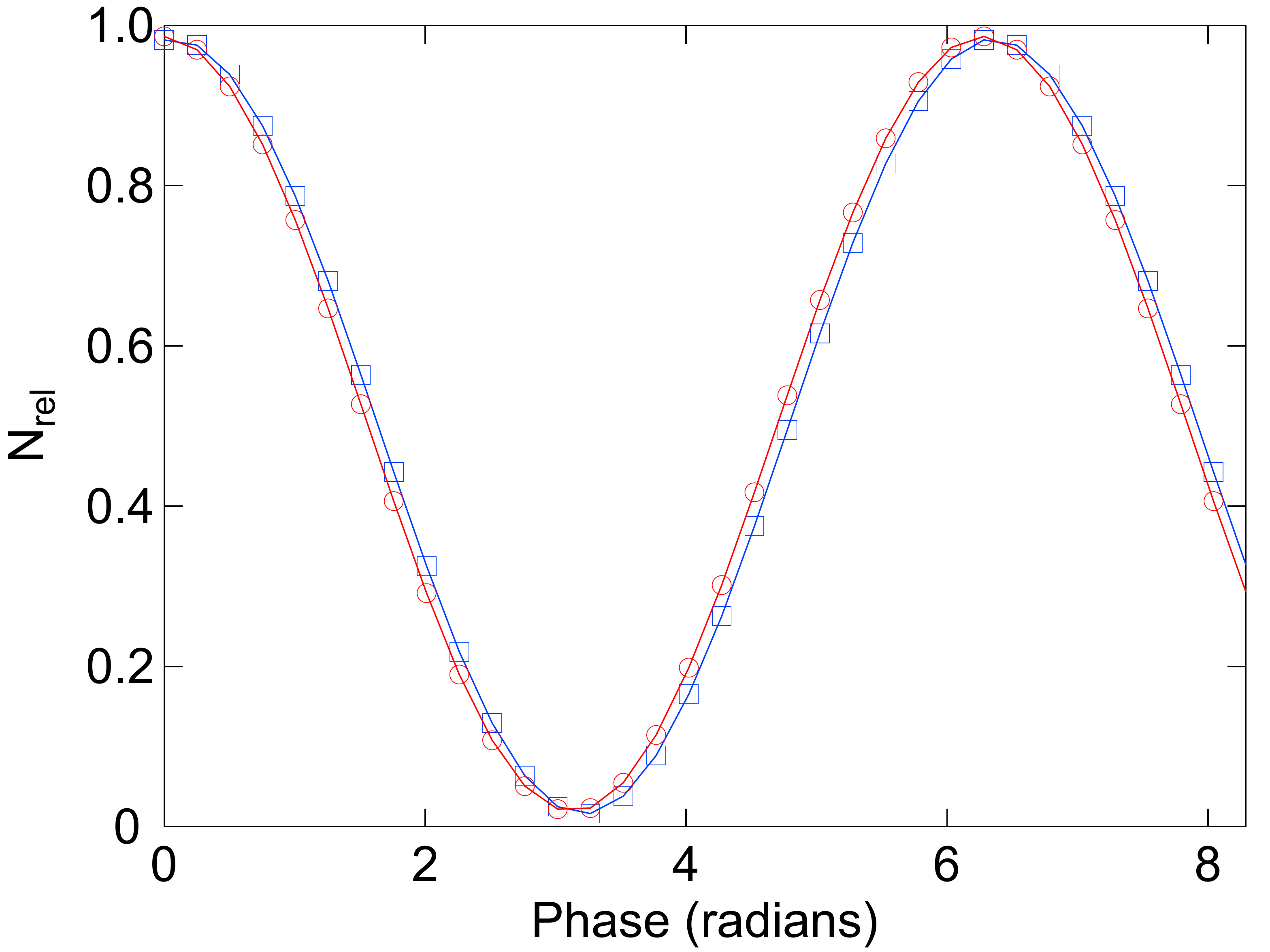}
 \caption{(Color online) Interference fringes for a BEC source obtained from the GP model for a $T=0.35$ $d.u.$, $2$ $\hbar k$ interferometer with $100$ $\%$ $\pi$ (cirlces) and $80$ $\%$ $\pi$ (squares) pulse.  This model agrees with the experimental results showing an increase in fringe contrast above the $\pi$ pulse efficiency for short $T$.  A small phase offset between the two curves is present.  Fig \ref{PhaseCompare}a corresponds to longer $T$ to accentuate the phase offset.}
 \label{Compare}
 \end{figure}

The multipath interference observed in the overlapping of BEC impurity states can be modeled numerically by reducing the 3D system to 1D.  The dimensionality reduction can be performed by writing an equivalent equation for the system in the dimension of interest \cite{schneider}.  A Gross-Pitaevskii (GP) model \cite{PhysRevA.69.051602,0953-4075-33-19-311} of the Bose-condensed Bragg interferometer in one dimension (co-linear with the Bragg optical lattice) is then given by

\begin{equation}
 \label{secondset}
\begin{array}{l} 
{\displaystyle
i\dot\phi_0=({\mathcal{L}}+
\frac{1}{2}x^2+Gx)\phi_0+\Omega\phi_n e^{ikx}}\\[7pt]
{\displaystyle
i\dot\phi_n=({\mathcal{L}}+Gx+\Delta)\phi_n+\Omega\phi_0 e^{-ikx}}\\[7pt]
\end{array}\end{equation}

where $\phi_0$ and $\phi_n$ are the GP wavefunctions for the $0$ and $n$th order momentum states, respectively, and ${\mathcal{L}}\equiv-\frac{1}{2}\frac{\partial^2}{\partial x^2}+U (\ |\phi_{0} |^2+|\phi_{n} |^2)$. Here $\Delta$ and $\Omega$ are the energy detuning between momentum states, and the Rabi frequency of the coupling, measured in units of the longitudinal trapping frequency  prior to release into the waveguide $\omega_{\parallel}$ ( of the $|F=1,m_F=1\rangle$ state), $U$ is the interaction coefficient and $G=\frac{mg}{\hbar\omega_{\parallel} }(\frac{\hbar}{m\omega_{\parallel} })^{1/2}\sin\theta$ is the dimensionless component of gravity along the horizontal guide. The wave functions, time, spatial coordinates, and interaction strengths are measured in the units of $(\hbar/m\omega_{\parallel})^{-1/4}$, $\omega_{\parallel} ^{-1}$, $(\hbar/m\omega_{\parallel} )^{1/2}$, and $(\hbar \omega_{\parallel} )^{-1}(\hbar/m\omega_{\parallel} )^{-1/2}$, respectively. The nonlinear interaction strength is derived by requiring that the 1D Thomas-Fermi chemical potential along the guide be equivalent to the full 3D case. In general the initial conditions are calculated using the relaxation method \cite{numerical} for the time independent solution, $\phi_0(x)$, in the potential $\frac{1}{2}x^2$, and solving the time dependent equations with only the potential $G=3$, essentially simulating free propagation along the guide with a milliradian tilt.  There are no free parameters in this model; $U=5.4\times10^{-3}$, $\Delta=-418$, and $k=29$ is used.  Fig (\ref{Compare}) shows the visibility curves for a $2$ $\hbar k$, $T=1$ ms interferometer with $100$ and $<80$ $\%$ $\pi$ pulses obtained from the above GP model.  The modeled interferometer corresponds, in cloud separation, to the experimental interferometer at $T=0.2$ $ms$.  

\begin{figure}
 \centering{}
 \subfigure{
 \includegraphics[width=.45\columnwidth]{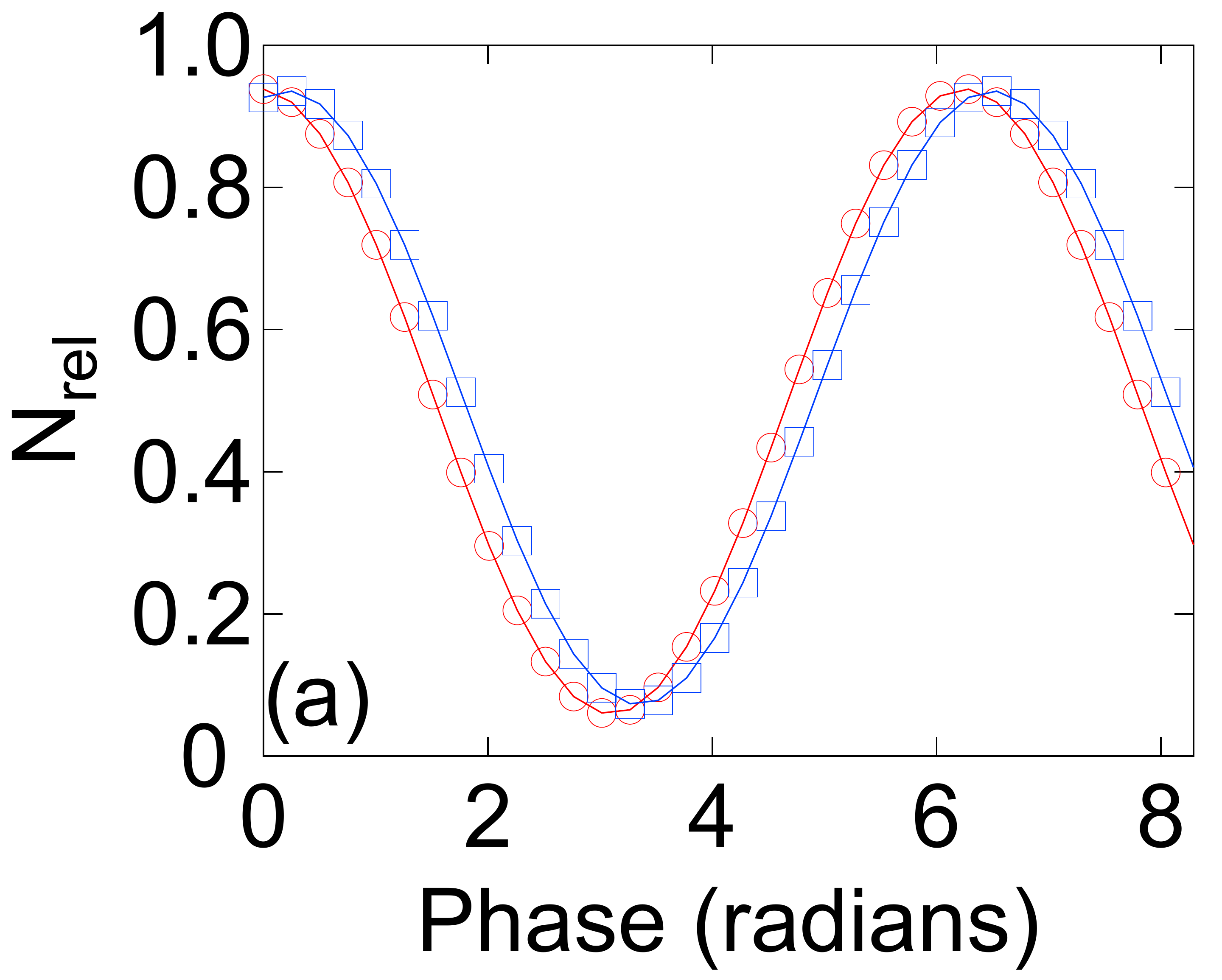}
 \label{TheoryOffset}}
 \hspace{-0.5 cm}
 \subfigure{
 \includegraphics[width=.45\columnwidth]{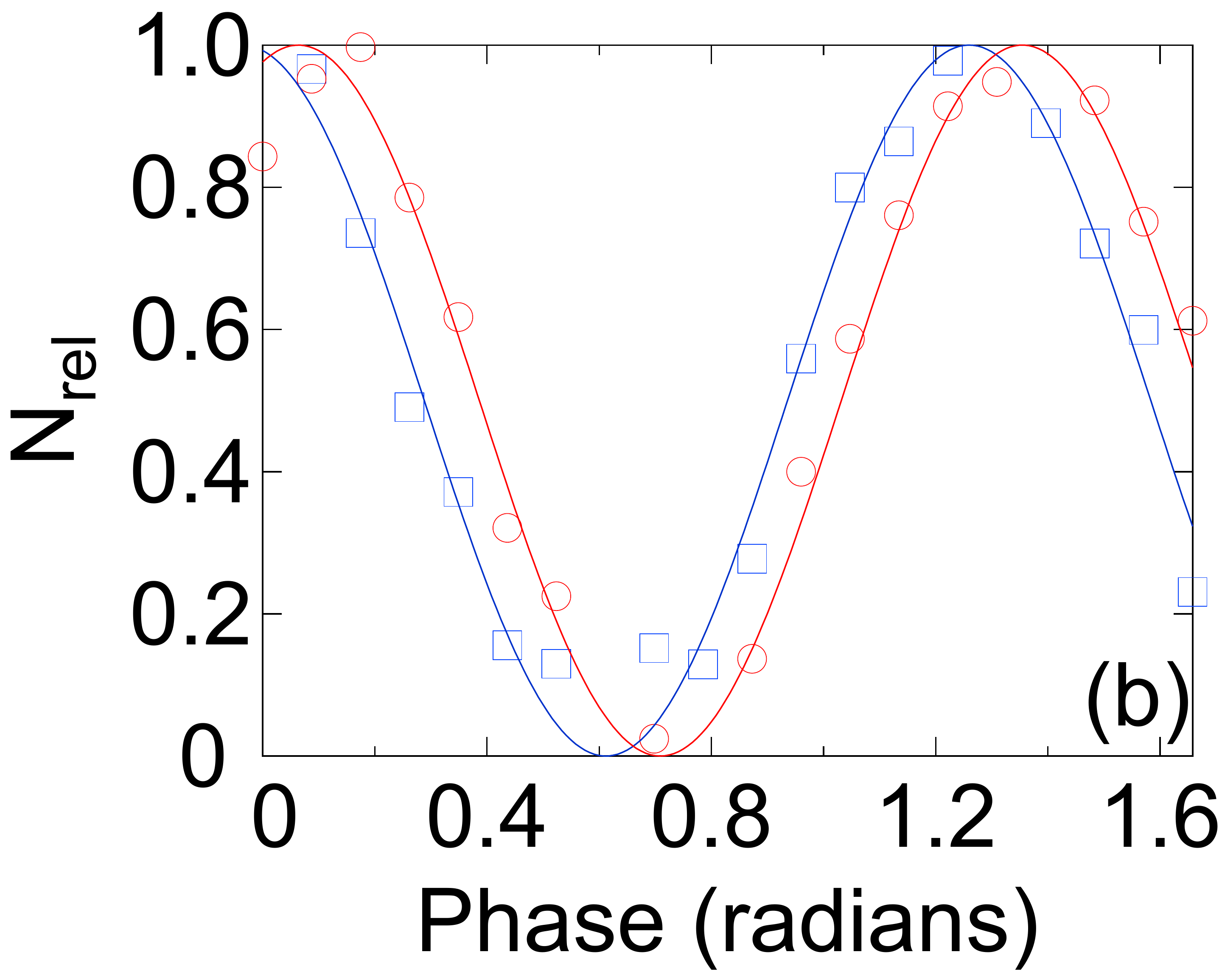}
 \label{CompareEXP}}
 \caption{ (Color online) (a) Interference fringes for a BEC source obtained from the GP model for a $T=1.15$ ms, $2$ $\hbar$k interferometer with $100$ $\%$ $\pi$ (circles) and $80$ $\%$ $\pi$ (squares) pulse. (b) Experimental interferometer fringes from a BEC source for a $T=1$ ms, $10$ $\hbar$k interferometer with $90$ $\%$ $\pi$ (circles) and $50$ $\%$ $\pi$ (square) pulse effeciencies.  These show the expected phase shifts induced by the interference from the velocity dependent phase of the impurity states.  The $50$ $\%$ $\pi$ pulse interferometer appears to be retarded in phase however it has undergone an integer number of period advances. The experimental curves have been normalized in $N_{rel}$.}
 \label{PhaseCompare}
 \end{figure}

\begin{figure}
 \centering{}
 \includegraphics[width=1\columnwidth]{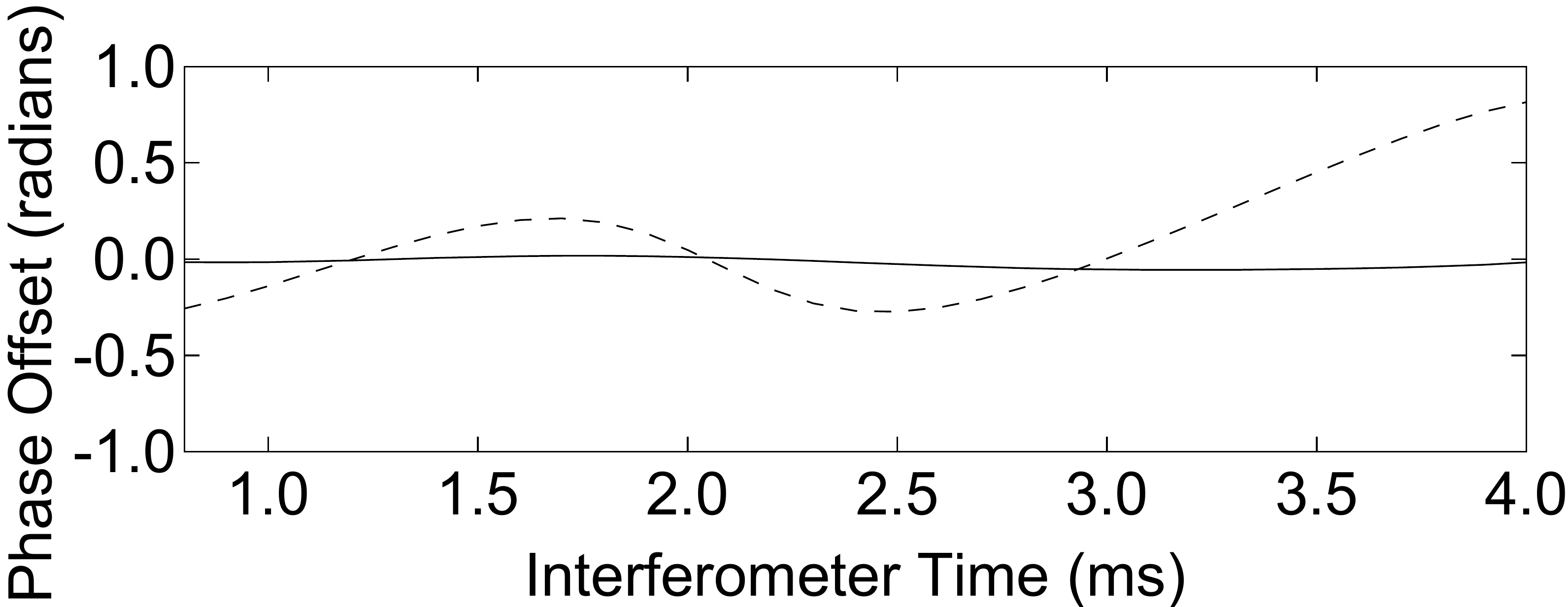}
 \caption{ The expected Phase offset as obtained from the GP model of a strongly interacting BEC source, $2$ $\hbar$k, interferometer with $100$ (solid line) and $80$ (dashed line) $\%$  efficient $\pi$ pulses as $T$ is increased.  The flat trend of the $100$ $\%$ $\pi$ efficient interferometer behaves as expected, from a thermal source with only minor fluctuations attributed to the strongly interacting BEC's mean field energy.  The $80$ $\%$ $\pi$ efficient shows oscillator behavior that obtains significant phase offset from the ideal system.}
 \label{Phaseoffsettheory}
 \end{figure}

Consistent with the experimental data, the GP simulation shows fringe visibility greater than the $\pi$ pulse efficiency, as seen in Fig \ref{Compare}.  This increase in visibility is due to the four path interference from the two main states and the two impurity states.  An added phase offset for the interferometer, in which impurities are present, is visible in Fig \ref{Compare} but accentuated in Fig \ref{PhaseCompare}a, where the interferometer time has been increased by a factor of three.   Fig \ref{PhaseCompare}b is experimental data showing the expected phase offset by comparing two $T=1$ ms interferometers with $90$ and $50$ $\%$ $\pi$ pulse efficiencies.  The curves have been normalized in $N_{rel}$ to illustrate the phase offset between the two interferometers.  This phase offset is non-trivially dependent on $T$, as the impurity states interfere in a Ramsey like interferometer configuration whose phase has both $T$ and $T^{2}$ dependent components \cite{PhysRevLett.67.177}.  Fig \ref{Phaseoffsettheory} shows the added phase accumulated by the interferometer when $T$ is increased as predicted by the GP model.  It is seen that a strongly interacting BEC source with a $100$ $\%$ $\pi$ pulse has similar phase shifts as expected from the thermal source, ie. $\approx 0$ added phase shift in $T$.  An $80$ $\%$ efficient $\pi$ pulse interferometer gives rise to an oscillatory phase offset in $T$ which will drastically effect the system's measured output.  For short time interferometers or any systems in which impurity states may be overlapped with the desired interferometer clouds this must be accounted for when processing the sensor's signal.  

\subsection{Interferometer: $T>4.5$ ms}
\begin{figure}
\centering{}
\subfigure{
\includegraphics[width=.45\columnwidth]{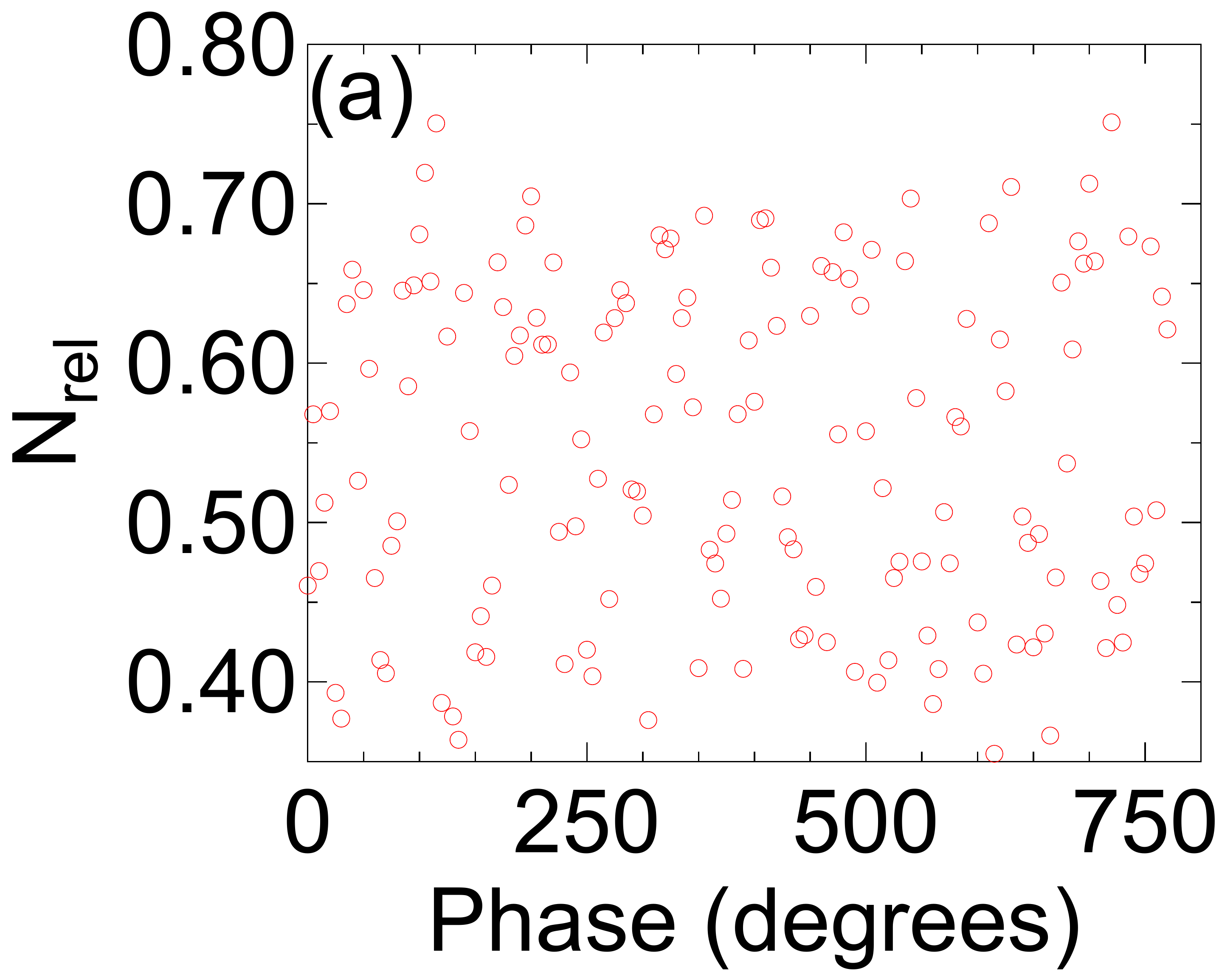}
\label{fringeBEC}}
\hspace{-0.4 cm}
\subfigure{
\includegraphics[width=.45\columnwidth]{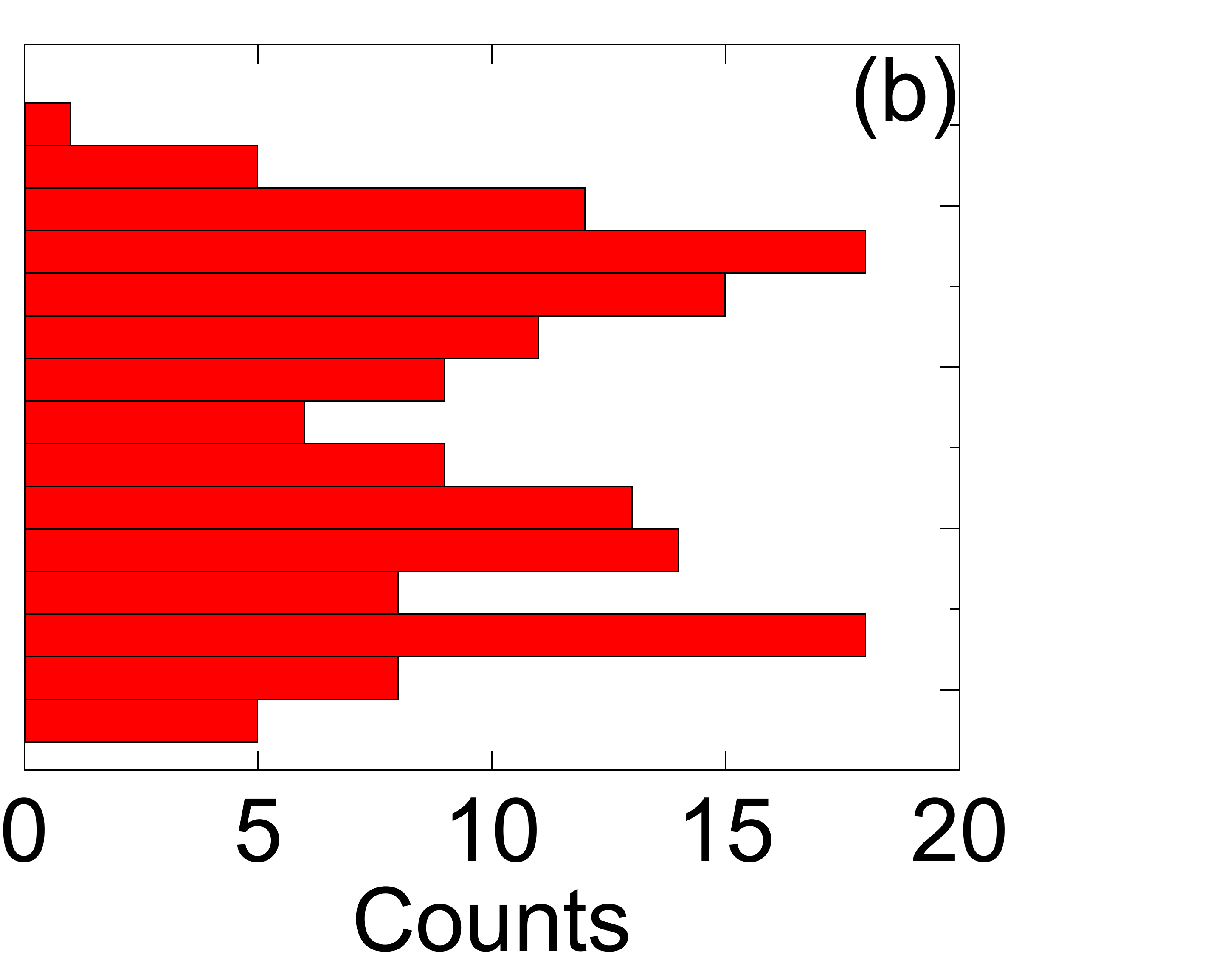}
\label{histoBEC}}
\quad
\subfigure{
\includegraphics[width=.45\linewidth]{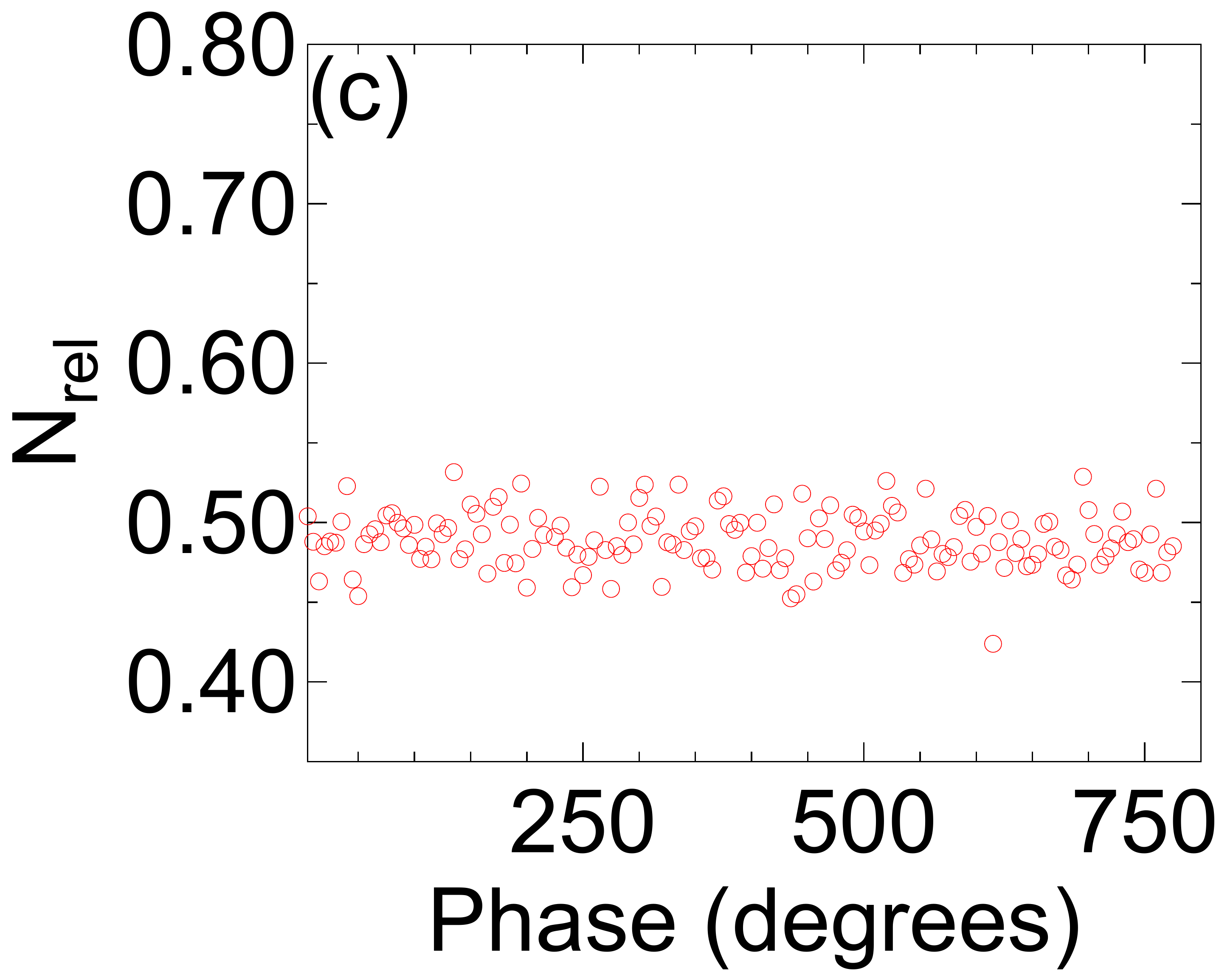}
\label{fringeTHERMAL}}
\hspace{-0.4 cm}
\subfigure{
\includegraphics[width=.45\linewidth]{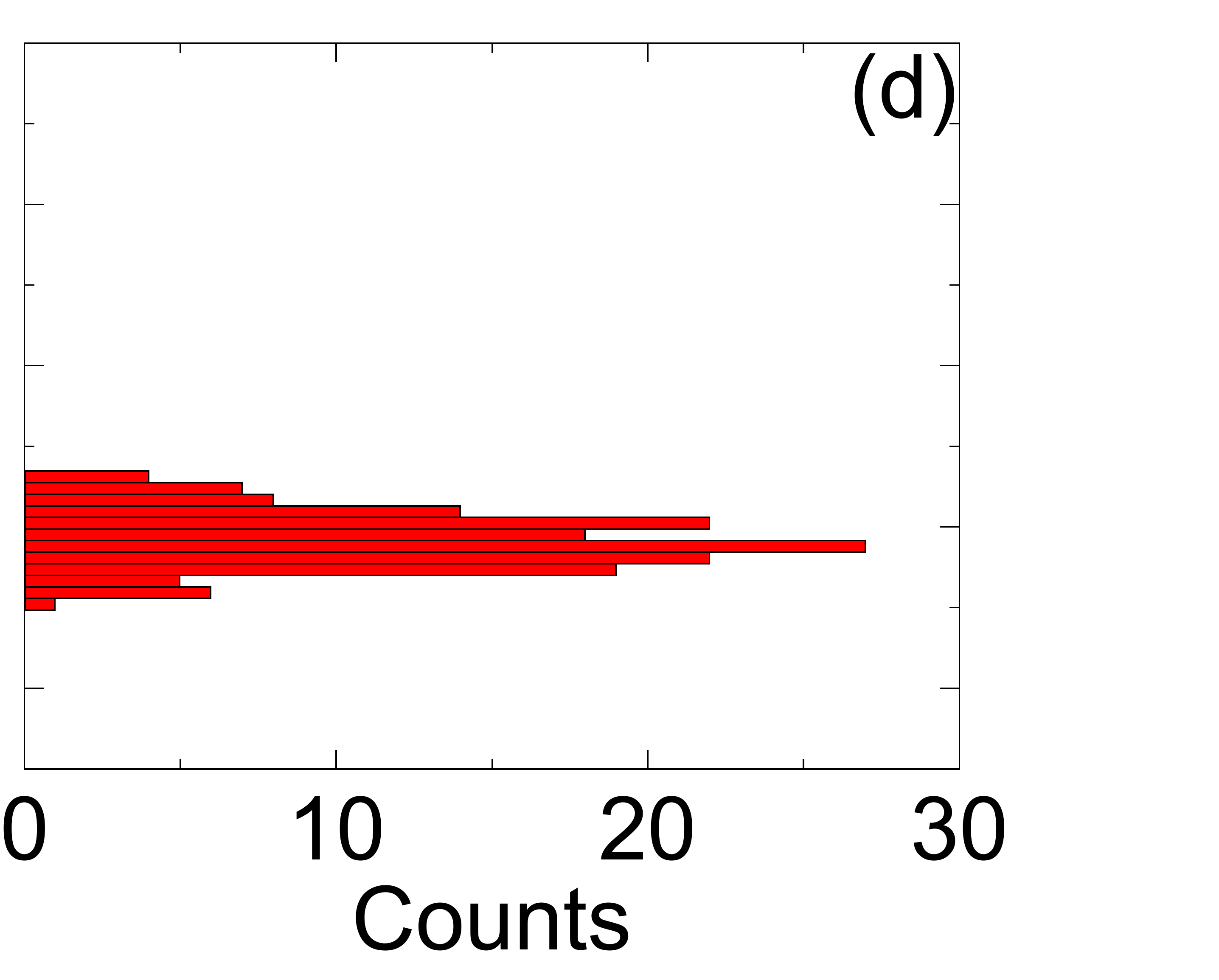}
\label{histoTHERMAL}}
\caption{(a) Interferometer output for a BEC source, $T=5.5$ ms, Bragg order$=5$, and $80$ $\%$ $\pi$ pulse with zero visibility.  (b) Histogram created from $15$ binned, evenly spaced, sections of the vertical axis of (a).  The bi-modal natural of this data is indicative of interference.  This data shows a contrast of $\approx 20$ $\%$.  (c) Interferometer output for a $\Delta p_{\perp }=0.41$ $\hbar k$ thermal source, $T=5.5$ ms, Bragg order$=5$, and $\approx 80$ $\%$ $\pi$ pulse with zero visibility.  (d)  Histogram created from $15$ binned, evenly spaced, sections of the vertical axis of (c).  The single peak nature of this histogram implies that no interference is present and therefore zero contrast.}
\label{FringeHisto}
\end{figure}

In order to investigate the role of BEC and thermal coherence length at longer $T$, fringe contrast at $T$ much greater than the extrapolated visibility zero crossing, from Fig \ref{visibility}, is measured for the BEC and the $\Delta p_{\perp}=0.41$ $\hbar$k thermal source, shown in Fig \ref{FringeHisto}.  To ensure no contribution from the impurity states, the data was taken at $T=5.5$ ms and $\approx 100$ $\%$ $\pi$ pulses were used.  It was calculated and experimentally confirmed that at this $T$, there was no spatial overlap between any impurity and the main interferometer states.  The direct comparison between contrast for the BEC and thermal source can be seen in Fig \ref{FringeHisto}a and \ref{FringeHisto}c.  Figures \ref{FringeHisto}b and \ref{FringeHisto}d illustrate the histograms associated with $15$ binned, evenly spaced sections of $N_{rel}$ from the data in Fig \ref{FringeHisto}a and \ref{FringeHisto}c.  The bunching of data points at two locations on either side of the mean in Fig \ref{FringeHisto}b is consistent with interference.  This is because a sinusoid with random  phase will generate a bi-modal histogram when binned according to this method \cite{Kasevich102hk, aip101}.  The single peak nature of the histogram in Fig \ref{FringeHisto}d indicates that there is no interference present at this $T$.  The BEC source maintains contrast of $\approx 20$ $\%$ and the thermal source shows noise and no measurable contrast.  The data does not show any measurable visibility as adequate vibration isolation has yet to be added to the system.  The impact of vibrations can be seen by calculating the required amplitude of a vibration which will wash out visibility.  At $T=5.5$ ms with a $10$ $\hbar$k inteferometer, a $1$ kHz vibration with an amplitude of $10^{-12}$ m  will give a one fringe phase shift.  The interference of the BEC source, at interferometer times well beyond where interference in thermal sources can be seen, is attributed to the fundamentally extended coherence length of the BEC, allowing for greater tolerance to spatial offset, induced from various noise sources, during the recombination pulse.  This maintained contrast is consistent with the prediction of the first order spatial coherence function, ${g_{1}(\Delta x)=\left(1-\frac{N_{c}}{N}\right) e^{-\pi\left( \frac{\Delta x}{\lambda_{T}}\right) ^{2}}+\frac{N_{c}}{N}}$, where $\Delta x$ is the spatial offset of the interfering clouds ($\Delta x<$cloud width), $N_{c}$ is the condensed atom number and $N$ is the total atom number \cite{PhysRevA.59.4595}.

\subsection{Conclusions}
A source cloud which is robust to both natural and systematic noise sources is desirable for atom interferometery due to the many imperfections in atom interferometer systems.   These imperfections include the inability to produce perfect $\pi$ and $\frac{\pi}{2}$ pulses, elastic scattering during cloud separation, and run to run classical noise sources which limit the ability to spatially overlap the interferometer states at long $T$.  It has been shown that atom interferometer contrast depends strongly upon the spatial coherence of the source.  A BEC source with spatial coherence limited only by the size of the condensate is the optimal source for such non ideal systems.  For all systems where impurity states are present, the BEC source will increase the contrast of the interferometer to beyond that which is achievable with a thermal source.  The overlapping of impurity states with the main interferometer clouds during the recombination pulse leads to a non trivial phase dependence in $T$.  This could have a detrimental effect in precision sensors operating very near on below the condensation point and lead to incorrectly mapping the measured phase to absolute signal.  This makes it critical to know the sources coherence length.  Furthermore, the coherence length of the BEC source makes it less sensitive to any spatial offsets at the recombination pulse.  This allows for contrast at $T$ greater than that which is possible with a thermal source, thereby increasing the sensitivity of the system.

\subsection{Future Directions}
Currently the interferometer system is being modified to produce a $^{85}$Rb condensate in which the interaction strengths may be modified via magnetic Feshbach resonances.  By moving to the non-interacting regime, scattering properties that lead to the creation of impurities when separating momentum states will be eliminated, allowing a pure BEC in the waveguide without degradation during the interferometer.  With the $^{85}$Rb condensate apparatus a direct comparison between free-space and guided interferometry will be made for both interacting and non-interacting cases.  The advantages of a BEC source demonstrated here make it important to implement and test such sources at the current limits of precision measurement.  

\subsection{Acknowledgments}
Carlos C. N. Kuhn would like to acknowledge financial support from CNPq (Conselho Nacional de Desenvolvimento Cientifico e Tecnologico).  John E. Debs would like to acknowledge financial support from the IC postdoctoral fellowship
program.  

\bibliographystyle{apsrev_v2}
\bibliography{Biblio}
\end{document}